\documentclass[useAMS,usenatbib]{mn2e}
\usepackage{graphicx,psfig,cite,epsf}  
\usepackage{times}
\usepackage{subfigure}
\usepackage{textcomp}
\usepackage{multirow}

\def\apjl{ApJ}
\def\apjl{ApJ}                % Astrophysical Journal, Letters
\def\apjs{ApJS}               % Astrophysical Journal, Supplement
\def\ao{Appl.~Opt.}           % Applied Optics
\def\aap{A\&A}                % Astronomy and Astrophysics
\def\inte{{\em INTEGRAL}}

\title[Study of the reflection spectrum of GX 3+1]{Study of the reflection spectrum of the accreting neutron star \\GX 3+1 using \textit{XMM-Newton} and \textit{INTEGRAL}}
\author[Pintore et al.]{\Large{Pintore F. $^1$, Di Salvo T. $^2$, Bozzo E.$^3$, Sanna A.$^1$, Burderi L.$^1$, D'A\`i A.$^2$, Riggio A.$^1$, Scarano F.$^1$, Iaria R.$^2$} \\
$^1$ Universit\`a degli Studi di Cagliari, Dipartimento di Fisica, SP Monserrato-Sestu, KM 0.7, 09042 Monserrato, Italy \\
$^2$ Dipartimento di Fisica e Chimica, Universit\'a di Palermo, via Archirafi 36 - 90123 Palermo, Italy \\
$^3$ ISDC - Science data center for Astrophysics, Ch. d'Ecogia 16, 1290, Versoix (Geneva), Switzerland \\ }

\begin{document}

\maketitle

\begin{abstract}
Broad emission features of abundant chemical elements, such as Iron, are commonly seen in the X-ray spectra of accreting compact objects and their studies can provide useful information about the geometry of the accretion processes.
In this work, we focus our attention on GX 3+1, a bright, persistent accreting low mass X-ray binary, classified as an atoll source. Its spectrum is well described by an accretion disc plus a stable comptonizing, optically thick corona which dominates the X-ray emission in the 0.3--20 keV energy band. In addition, four broad emission lines are found and we associate them with reflection of hard photons from the inner regions of the accretion disc where doppler and relativistic effects are important. We used self-consistent reflection models to fit the spectra of the 2010 \textit{XMM-Newton} observation and the stacking of the whole datasets of 2010 \textit{INTEGRAL} observations. We conclude that the spectra are consistent with reflection produced at $\sim10$ gravitational radii by an accretion disc with an ionization parameter of $\xi\sim$ 600 erg cm s$^{-1}$ and viewed under an inclination angle of the system of $\sim35$\textdegree. Furthermore, we detected for the first time for GX 3+1, the presence of a powerlaw component dominant at energies higher than 20 keV, possibly associated with an optically thin component of non-thermal electrons. 

\end{abstract}


\section{Introduction}
\label{intro}

Broad, and possibly asymmetrical, emission features are commonly found in the X-ray spectra of accreting Galactic black holes (BHs; \citealt[e.g.][]{miller02a}), supermassive BHs in Active Galactic Nuclei (AGN; \citealt{tanaka95,fabian02}) and neutron star systems (NS; e.g. \citealt{bhattacharyya07,dai09,disalvo09}). The main observed reflection features can be ascribed to transitions of the K-shell of iron in the energy range 6.4--7.0 keV, according to its different state of ionizations. In addition, broad features associated to lighter chemical elements have been found in the spectra of a number of accreting sources \citep[e.g.][]{disalvo09,iaria09,dai10,egron13}.
It has been proposed that the broadening of such features can be produced by either i) relativistic reflection of hard photons from the surface of the accretion disc \citep[e.g.][]{fabian89,reynolds03, fabian05, matt06} or ii) Compton broadening if the line is emitted in the inner parts of a moderately, optically thick accretion disc corona (ADC), created by the evaporation of outer layers of the disc \citep{white82a,kallman89, vrtilek93}. iii) Another possibility is Compton down-scattering due to a narrow wind shell ejected at mildly relativistic velocities at some disc radii where the local radiation force overcomes the local disc gravity \citep{titarchuk09}. We note also that \citet{ng10} proposed that in many cases the asymmetrical shape of the spectral features seen in \textit{XMM-Newton}/EPIC-pn data may be due to a wrong estimation of pile-up, although, on the contrary, \citet{miller10} showed that modest level of pile-up should not significantly affect the spectral analysis.
However, the relativistic reflection scenario described above appears the more suitable explanation in many cases (e.g. Di Salvo et al. 2015, in press). 

The broad lines produced by relativistic reflection of hard photons from the disc are usually double-peaked due to the motion of matter of the disc. Furthermore, the blueshifted peak is also relativistically boosted as a consequence of the matter in the accretion disc approaching the observer at speed values of the order a few tenths of the speed of light when it is at a few gravitational radii from the compact object. The broad lines are then gravitationally redshifted when produced close to the compact object. The combination of such effects highly distorts the spectral shape of the line \citep{fabian89}.
These effects are observed in many broad lines of abundant chemical elements as Iron. In addition to the broad Iron emission feature, commonly detected at 6.4--7.0 keV, an absorption edge at $\sim 8-9$ keV is usually observed.

Moreover, a large fraction of hard X-ray photons is reflected by the accretion disc producing a reflection continuum, due to the direct Compton scattering, commonly known as ``Compton bump'' which peaks at around 20--30 keV \citep[e.g.][]{george91}.
Even though often described independently, the discrete emission features and the reflection continuum are closely related. Self-consistent models can instead investigate the physics of the reflection processes taking into account both spectral components at the same time (\citealt[][]{ross99,ross07,reis08,cackett10,dai10,egron13,disalvo15}). 
The detailed study of the reflection component plays therefore a fundamental role in the investigation of the properties of the compact objects (BHs and NSs) and the physical mechanisms of the accreting material around them. 

In this work we investigate the spectral properties of the Galactic source GX 3+1 \citep{bowyer65}, which is a bright, persistent accreting low mass x-ray binary (LMXB). A number of type I X-ray bursts, likely due to burning helium, has been detected, with decay timescales ranging from tens of seconds up to hours \citep[e.g.][]{makishima83,denHartog03,kuulkers00, kuulkers02,chenevez06}. The presence of thermonuclear bursts suggest that the compact object hosted in GX 3+1 is an accreting NS. The estimated upper limit to its distance is $\sim 6.1$ kpc \citep{kuulkers00}. A probable quasi periodic oscillation (QPO) with a frequency of 8 Hz has been detected during an EXOSAT observation \citep{lewin87}.
Based on its variability and spectral properties, GX 3+1 was classified as an atoll source \citep{hasinger89}. \citet{vandenberg14} identified a counterpart of GX 3+1 in the near-infrared (NIR) with a Ks=15.8 +/- 0.1 mag suggesting that the donor star is not a late-type giant and that the NIR emission is dominated by the heated outer accretion disc. Spectral analyses of the source showed that its X-ray spectrum can be described by a model comprising of a soft blackbody component, most likely associated with the accretion disc, and a comptonizing component, produced by an optically thick electron population located close to the NS boundary layer \citep{mainardi10,piraino12,seifina12}. \citet{seifina12}, using all the Rossi X-ray Timing Explorer (RXTE) and BeppoSAX data, showed that the comptonizing component is quite stable suggesting that the energy release in the transition layer between the accretion disc and the NS dominates over the energy release in the accretion disc. A number of broad emission lines were also reported by \citet{piraino12} and associated with the K${\alpha}$ transitions of Ar, Ca and Fe. These features have been ascribed to relativistic reflection and the inferred inner disc radius of $\sim$30 gravitational radii indicated a truncated accretion disc \citep{piraino12}.

In next section, we will carry out a deep spectral analysis of the broadband spectrum of the stacking of the whole 2010 \textit{INTEGRAL} observations and the 2010 \textit{XMM-Newton} observation. In particular, we will study the properties of the reflection component, as shown in \citet{piraino12}, and extending it with the use of self-consistent models. The paper is structured as follows: in Section~\ref{data_reduction} we summarize the data selection and reduction procedures, in Section~\ref{analysis} we present the spectral results of GX 3+1 and discuss them in Section~\ref{discussion}.

\section{Data Reduction}
\label{data_reduction}

The datasets considered here comprised the \textit{XMM-Newton} observation (Obs.ID. 0655330201), taken in Timing mode on September 1st, 2010, combined with the whole available 2010 \textit{INTEGRAL} observations. 

\textit{XMM-Newton}/EPIC-pn data were reduced using the calibration files (at the date of April 25th, 2014) and the Science Analysis Software (SAS) v. 13.5.0. At high count rates, X-ray loading and charge transfer inefficiency (CTI) effects have to be considered because they modify the spectral shape and produce an energy shift on the spectral features (e.g. Guainazzi et al. 2008, XMM-CAL-SRN-248\footnote{http://xmm2.esac.esa.int/docs/documents/CAL-SRN-0248-1-0.ps.gz}). We corrected these effects by adopting the rate dependent pulse height amplitude corrections (RDPHA) which have been shown to provide a better accuracy of the energy scale calibration \citep{pintore14b}. We extracted the spectra from events with {\sc pattern}$\leq 4$ (which allows for single and double pixel events) and we set `{\sc flag}=0', retaining events optimally calibrated for spectral analysis. Source and background spectra were then extracted selecting the ranges RAWX=[31:43] and RAWX=[3:5], respectively. EPIC-pn spectra were rebinned in energy with an oversample of 7 using the \textit{specgroup} task. 
We looked for pile-up in the EPIC-pn data by comparing the source spectrum obtained considering all the columns of the CCD in the range RAWX=[31:43] and the spectra obtained excising one, three and five central brightest columns of the aforementioned RAWX range. We found that pile-up was only marginally important and it could be clearly mitigated removing only the brightest central column of the EPIC-pn CCD. Hereafter we will refer, if not specified, to spectral analysis without the central column.

We extracted first and second order RGS spectra using the standard \textit{rgsproc} task, filtered for periods of high background. After checking for spectral consistency of RGS 1 and 2 in both first and second order, we stacked together the RGS1 and RGS2 data adopting the tool \textit{rgscombine} which combines, separately, first and second order spectra. We did not find any feature in the RGS datasets, therefore we grouped the two final spectra with a minimum of 100 counts per energy channel. Since the background, in both orders, was not negligible for energies lower than 0.9 keV, we analyzed the RGS spectra in the range 0.9--2.0 keV. Even though the mean count rate is at least a factor of 3 lower (RGS1 $\sim$3.4 and RGS2 $\sim$1.7 cs s$^{-1}$) than the nominal RGS pile-up threshold, we investigated our data looking up for pile-up effects and we found that pile-up does not significantly affect our data because we verified that first and second order spectra are in complete agreement with each other. This allows us to confirm that pile-up problems are negligible for this source.

For the scope of the present paper, we also considered all the available \inte\ data collected in 2010 in the direction of GX\,3+1. The dataset included ``science windows'' (SCWs), i.e. the \inte\ pointings with typical durations of $\sim$2 ks each, from satellite revolution 895 to 981. All data were analyzed by using version 10.0 of the OSA software distributed by the ISDC \citep{courvoisier03}. We used data from IBIS/ISGRI \citep[17--80 keV,][]{lebrun03,ubertini03} in which the source was located to within 12\textdegree$ $  from the center of the instrument field of view (FoV), thus avoiding problems with calibration uncertainties. We also extracted JEM-X1 and JEM-X2 \citep[][]{lund03} data from the same set of SCWs. The ISGRI (JEM-X) response matrices were rebinned in order to have 37 (32) energy bins spanning the range 20--180 keV (3--35 keV) for all spectra. This optimized the signal-to-noise (S/N) of the data. 

We fitted simultaneously the \textit{XMM-Newton} (RGS and EPIC-pn) and \textit{INTEGRAL} (JEM-X1, JEM-X2 and INTEGRAL/ISGRI) spectra, using {\sc xspec} V. 12.8.2 \citep{arnaud96}. We selected the energy ranges 0.9--2.0 keV, 2.4--10.0 keV, 10.0--20.0 keV and 20.0--50.0 keV, for RGS, EPIC-pn, JEM-X1, JEM-X2 and ISGRI, respectively. We ignored the EPIC-pn channels at energies lower than 2.4 keV as we found a soft excess which was not reconciled with the RGS data, suggesting that the EPIC-pn calibrations are still uncertain below this energy (internal \textit{XMM-Newton} report CAL-TN-0083\footnote{http://xmm2.esac.esa.int/docs/documents/CAL-TN-0083.pdf}; see also \citealt{piraino12}).

We introduced a multiplicative constant model in each fit in order to take into account the diverse flux calibrations of the considered instruments and possible differences on the flux caused by non simultaneity of \textit{XMM-Newton} and \textit{INTEGRAL} observations. We fixed to 1 the constant for the EPIC-pn spectrum and allowed the other constants to vary. In general, the range of this parameter is consistent within 10$\%$ for the RGS instruments and does not vary more than $\sim20\%$ for the INTEGRAL/ISGRI and JEM-X, in comparison with the EPIC-pn constant.

\section{Spectral analysis}
\label{analysis}

The aim of the spectral analysis presented here is to provide a characterization of the X-ray emission from GX 3+1 in a broadband energy range. We are particularly interested in the reflection component whose properties can be accurately described only when the overall continuum is well constrained. Previous works \citep[e.g.][]{seifina12} have shown that the high energy component of GX 3+1 is weakly variable. 

\begin{figure}
\center
\includegraphics[height=8.4cm,width=6.9cm,angle=270]{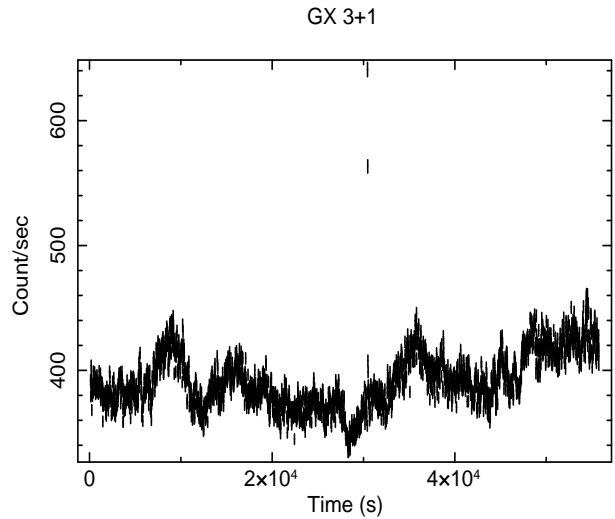}
\caption{\textit{XMM-Newton}/EPIC-pn lightcurve in the 0.3--10 keV energy range, corrected for pile-up excising the brightest central column and sampled with bins of 20 seconds. A type-I burst and clear time variability are observed.}
\label{tot_lc.eps}
\end{figure}

We analyse the 2010 \textit{XMM-Newton} observation with a net exposure time of 55 ks. In Figure~\ref{tot_lc.eps}, we show the total light curve in the 0.3--10 keV energy band: during the exposure, a type I burst, which characterizes the temporal properties of the source, was observed and GX 3+1 presented also an unambiguous evidence of flux variability on timescales of $\sim5$ ks. To test the contribution of the burst in the spectral parameters, we fit RGS and EPIC-pn spectra (with and without the burst time interval) adopting a self-consistent model, as showed in section~\ref{selfcons}, and we find that the X-ray burst does not affect our spectral parameters.
For the temporal variability, we instead investigate if such variability could also drive spectral changes. {To test this, we follow two alternative approaches based both on temporal and flux variability: in the first one, we split the observation in 11 temporal segments of 5 ks each; while, in the second one, in order to further improve the signal-to-noise ratio, we divide the observation in three flux binned segments (low flux ($<370$ cs s$^{-1}$), medium flux (between $370$ cs s$^{-1}$ and 400 cs s$^{-1}$) and high flux ($>400$ cs s$^{-1}$)) and with a similar amount of collected counts.} We extract an EPIC-pn spectrum, pile-up corrected, for each segment and we fit the obtained spectra in the 2.4--10 keV energy range, where the calibration of the instrument is more accurate. We adopt a simple comptonization model based on an {\sc nthcomp} \citep{zdiarski96} plus a relativistic line ({\sc diskline}, \citealt{fabian89}; see below for more details). We find that all the spectral parameters inferred by the spectral analysis are consistent within the uncertainties (at 3$\sigma$ confidence level). 
We thus report in the following only on the average spectrum of the whole \textit{XMM-Newton} observation. 

\begin{figure}
\center
\includegraphics[height=8.4cm,width=6.9cm,angle=270]{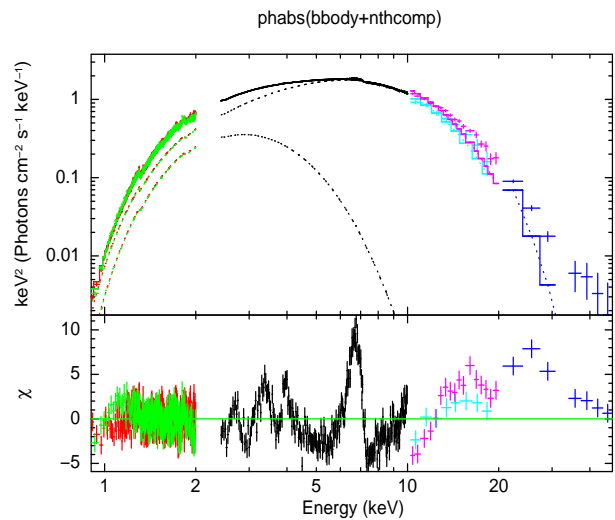}
\caption{Unfolded $E^2f(E)$ \textit{XMM-Newton}/EPIC-pn (\textit{black}), \textit{XMM-Newton}/RGS spectra (\textit{red} and \textit{green}), \textit{INTEGRAL}/JEM-X1 and X2 (\textit{cyan} and \textit{magenta}), and \textit{INTEGRAL}/ISGRI (\textit{blue}). The dashed curves represent the {\sc bbody} and the {\sc nthcomp} components (see text). For display purpose only, RGS data have been rebinned with a minimum of 20$\sigma$ and maximum number of energy channels of 20.}
\label{GX_3+1_continuum}
\end{figure}

\begin{figure*}
\center
\subfigure{\includegraphics[height=7.8cm,width=6.9cm,angle=270]{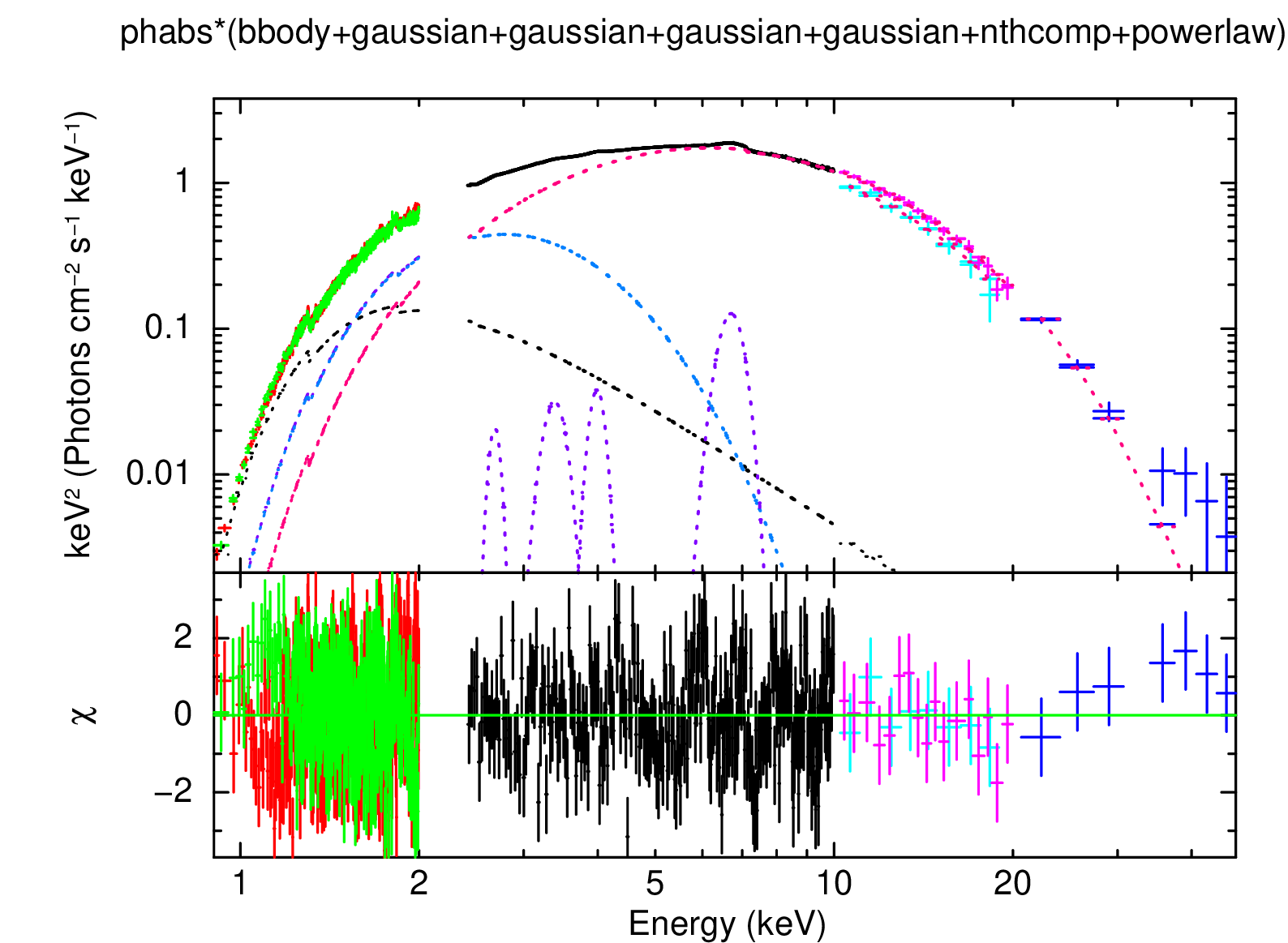}}
\subfigure{\includegraphics[height=7.8cm,width=6.9cm,angle=270]{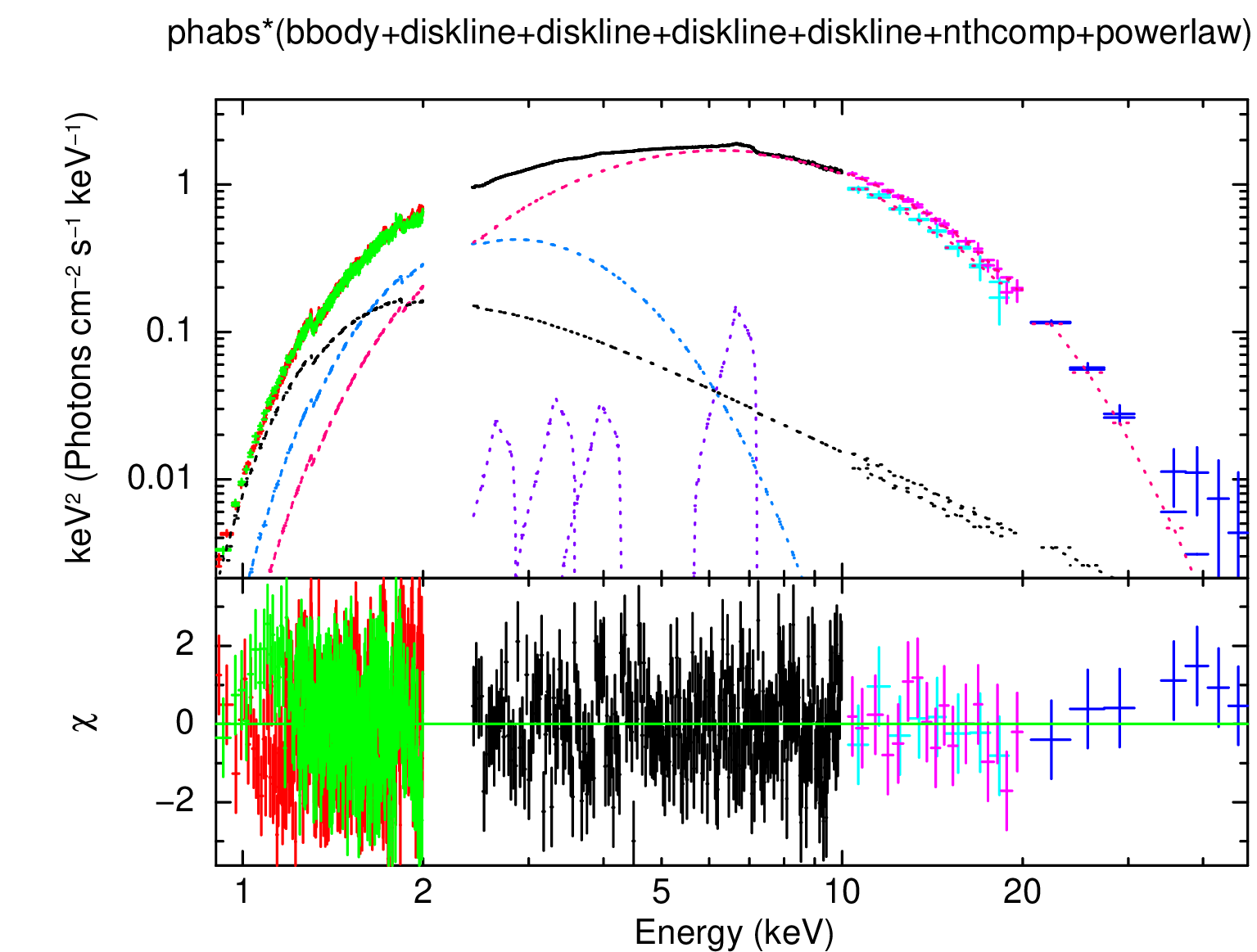}}
\subfigure{\includegraphics[height=7.8cm,width=6.9cm,angle=270]{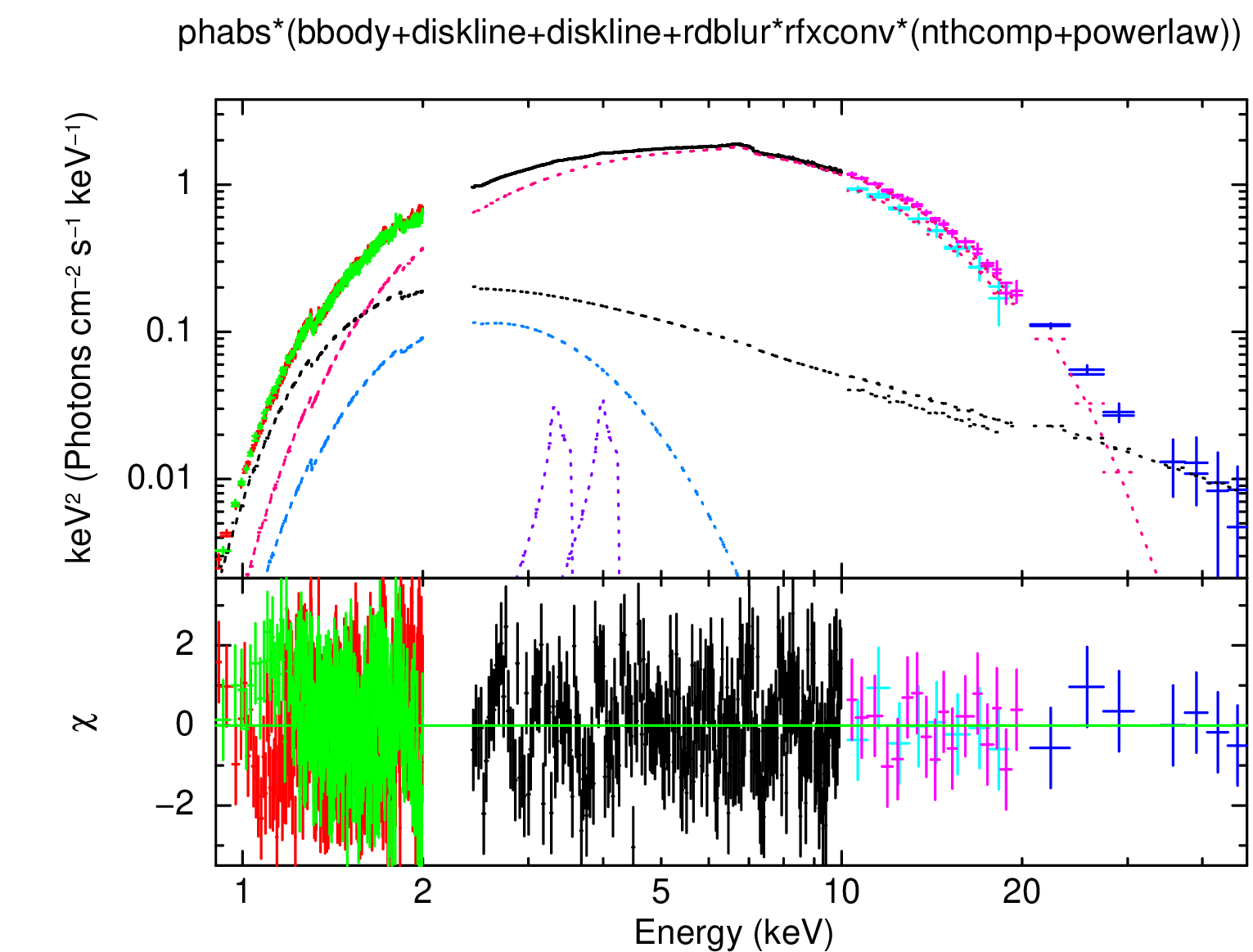}}
\subfigure{\includegraphics[height=7.8cm,width=6.9cm,angle=270]{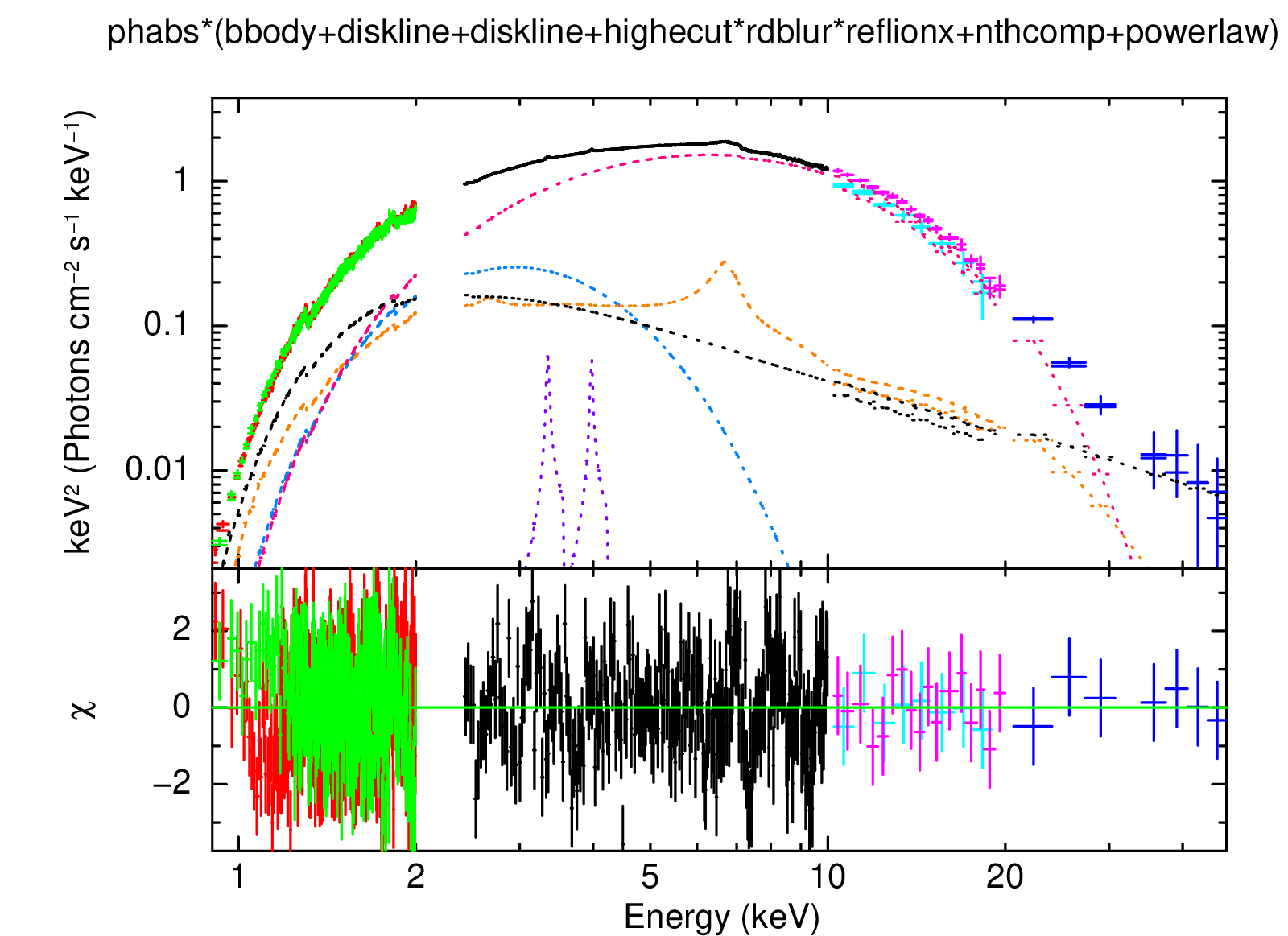}}
\caption{Unfolded $E^2f(E)$ \textit{XMM-Newton}/EPIC-pn (\textit{black}), \textit{XMM-Newton}/RGS spectra (\textit{red}, first order, and \textit{green}, secondo order), \textit{INTEGRAL}/JEM-X1 and X2 (\textit{cyan} and \textit{magenta})  and \textit{INTEGRAL}/ISGRI (\textit{blue}). The top panel shows the best fit with the {\sc gaussian} (top-left) or {\sc diskline} (top-right) models; the bottom panel shows the best fit with the {\sc rfxconv} (bottom-left) or {\sc reflionx} (bottom-right) self-consistent models. The dashed curves represent the {\sc bbody} (\textit{light blue}), the {\sc nthcomp} (\textit{light red}), the {\sc powerlaw}  (\textit{dark grey}), the {\sc gaussian} or {\sc diskline} (\textit{purple}) and the self-consistent (\textit{orange}) components. For display purpose only, RGS data have been rebinned with a minimum of 20$\sigma$ and maximum number of energy channels of 20.}
\label{GX_3+1_spectra}
\end{figure*}

We note that the Monitor of All-sky X-ray Image (MAXI) telescope observed GX 3+1 in a quiete stationary flux level in the 10--20 keV energy band during the entire 2010, with average variability lower than a factor of 2. Therefore, we also decide to stack all the available 2010 INTEGRAL observations of GX 3+1 in order to improve the signal-to-noise ratio in the band 10-50 keV. In addition, this choice allows us to collect very high quality data for a period which is reasonably close to the \textit{XMM-Newton} observation.
However, although the flux variability was limited during the 2010, we verify that also the spectral variability was negligible during the year: we compare the INTEGRAL spectra (both ISGRI and JEM-X instruments) obtained by stacking the data of the whole year and those obtained by stacking the data of about 100 days centered at the \textit{XMM-Newton} observation. We select 100 days as a test in order to have a good quality counting statistics. No significant spectral variability is detected.
The total exposure times of the INTEGRAL instruments is then 700 ks, 64 ks and 390 ks for ISGRI, JEM-X1 and JEM-X2, respectively. 
\newline
\newline
We initially investigate the spectral properties of GX 3+1 adopting a model consisting of an absorbed ({\sc phabs} in {\sc xspec}) blackbody ({\sc bbody}) plus a comptonizing component ({\sc nthcomp}), as it was shown in previous works to be a good representation of the source X-ray continuum \citep[e.g.][]{mainardi10,seifina12, piraino12}.
\begin{table*}
\footnotesize
\begin{center}
\caption{Best fit spectral parameters obtained with the absorbed continuum {\sc bbody+nthcomp+powerlaw} model. Column 1, 2, 3 and 4 refer to best fit models where the reflection component is described by {\sc gaussian, diskline, rfxconv} or {\sc reflionx} models, respectively. Errors are at 90$\%$ for each parameter.}
\label{table_continuum_comptb}
\scalebox{0.9}{\begin{minipage}{13.0cm}
\begin{tabular}{llllll}
\hline
Model & Component &\multicolumn{1}{c}{(1)} & \multicolumn{1}{c}{(2)} & \multicolumn{1}{c}{(3)} & \multicolumn{1}{c}{(4)}  \\
\\
{\sc phabs} 	    &N$_H$ (10$^{22}$ cm$^{-2}$)$^a$ &$2.19^{+0.08}_{-0.09}$ 	& $2.1^{+0.1}_{-0.1}$ 	&$2.01^{+0.03}_{-0.04}$ & $2.02^{+0.03}_{-0.04}$ 		 \\
\\
{\sc bbody} 	    &kT$_{bb}$ (keV)$^b$			    & $0.54^{+0.02}_{-0.02}$	& $0.57^{+0.02}_{-0.03}$ 	& $0.50^{+0.01}_{-0.03}$ & $0.60^{+0.01}_{-0.01}$  \\
& Norm. ($10^{-3}$)$^c$ 						    & $20.2^{+0.1}_{-0.2}$ 	& $17.9^{+0.2}_{-0.3}$ 	& $5.5^{+0.9}_{-0.9}$ 	& $10.2^{+0.1}_{-0.1}$		 \\
\\
{\sc nthcomp} 	     & kT$_{e}$ (keV)$^d$			    & $3.0^{+0.6}_{-0.3}$ 	& $3.1^{+0.8}_{-0.5}$	& $2.25^{+0.04}_{-0.04}$ 	& $2.21^{+0.04}_{-0.04}$	 \\
&$\Gamma$$^e$	     				    		    & $2.7^{+0.2}_{-0.3}$ 	& $2.9^{+0.3}_{-0.5}$	& $1.77^{+0.03}_{-0.02}$ 	& $1.80^{+0.05}_{-0.05}$	 \\
&kT$_{0}$$^f$	     							    & $1.2^{+0.07}_{-0.07}$ 	& $1.26^{+0.07}_{-0.08}$ 	& $0.69^{+0.06}_{-0.2}$    & $0.92^{+0.10}_{-0.08}$ 	 \\
& Norm. ($10^{-2}$)$^g$ 						    & $10.6^{+0.1}_{-0.1}$ 	& $10^{+1}_{-1}$ 	& $21^{+2}_{-2}$ 	& $12^{+2}_{-2}$ 		  \\
\\
{\sc powerlaw} &$\Gamma$$^e$	     			    & $4.7^{+0.4}_{-0.5}$ 	& $4.0^{+0.6}_{-0.7}$	& $3.36^{+0.10}_{-0.05}$ 	& $3.3^{+0.3}_{-0.1}$	 \\
& Norm. $^g$ 						    		    & $2.1^{+0.4}_{-0.2}$ 	& $1.5^{+0.7}_{-0.4}$ 	& $1.2^{+0.1}_{-0.2}$	& $0.9^{+0.1}_{-0.1}$ 		  \\
\\
{\sc gaussian} 	     & Energy (keV)$^h$			    & 2.68$_{-0.02}^{+0.03}$ 	& 		-			& 		-			& 	-			 		 \\
& {$\sigma$} $^i$ 							    & 0.06$_{-0.04}^{+0.05}$ 	& 		-			& 		-			& 	-					  \\
& Norm. ($10^{-3}$)$^j$ 						    & 0.7$_{-0.2}^{+0.5}$	& 		-			& 		-		 	& 	-					  \\
\\
{\sc gaussian} 	     & Energy (keV)$^h$			    & 3.37$_{-0.02}^{+0.02}$ 	& 		-			& 		-			& 	-		 			\\
& {$\sigma$} $^i$ 							    & 0.14$_{-0.03}^{+0.03}$	& 		-			& 		-		 	& 	-	  				\\
& Norm. ($10^{-3}$)$^j$ 						    & 1.3$_{-0.3}^{+0.4}$	& 		-			& 		-			& 	-	  				\\
\\
{\sc gaussian} 	     & Energy (keV)$^h$			    & 3.98$_{-0.02}^{+0.02}$ 	& 		-			& 		-			& 	-	 				\\
& {$\sigma$} $^i$ 							    & 0.12$_{-0.04}^{+0.05}$	& 		-			& 		-		 	& 	-	  				\\
& Norm. ($10^{-3}$)$^j$ 						    & 0.8$_{-0.2}^{+0.3}$ 	& 		-			& 		-			& 	-	  				\\
\\
{\sc gaussian} 	     & Energy (keV)$^h$			    & 6.68$_{-0.02}^{+0.02}$ 	& 		-			& 		-		 	& 	-					 \\
& {$\sigma$} $^i$ 							    & 0.32$_{-0.02}^{+0.02}$	& 		-			& 		-			& 	-	  				\\
			     & Norm. ($10^{-3}$)$^j$ 		    & 2.4$_{-0.2}^{+0.2}$	& 		-			& 		-		 	& 	-	  				\\
\\
{\sc diskline} 	     & Energy (keV)$^h$			   & 	-					& 2.64$_{-0.02}^{+0.02}$ 	& 		-		 	& 	-					 \\
		             & Norm. ($10^{-3}$)$^j$ 		   & 	-			 		& 1.6$_{-0.4}^{+0.4}$	& 		-		 	& 	-					  \\
\\
{\sc diskline}          & Energy (keV)$^h$ 			   & 	-				  	& 3.32$_{-0.02}^{+0.02}$ 	& 3.32$_{-0.02}^{+0.02}$  & 3.35$_{-0.02}^{+0.02}$ 		 \\
		             & Norm. ($10^{-3}$)$^j$ 		   & 	-					& 1.5$_{-0.2}^{+0.2}$ 	& 1.2$_{-0.2}^{+0.2}$	& 0.8$_{-0.1}^{+0.2}$ 		 \\
\\
{\sc diskline}          & Energy (keV)$^h$ 			   & 	-				  	& 3.95$_{-0.02}^{+0.02}$	& 3.98$_{-0.02}^{+0.02}$  & 3.97$_{-0.02}^{+0.02}$ 		 \\
		             & Norm. ($10^{-3}$)$^j$ 		   & 	-			  		& 1.1$_{-0.2}^{+0.2}$	& 0.9$_{-0.2}^{+0.2}$ 	& 0.6$_{-0.1}^{+0.1}$ 		 \\
\\
{\sc diskline}          & Energy (keV)$^h$ 			   & 	-				  	& 6.65$_{-0.02}^{+0.02}$ 	& 		-			& 	-					 \\
		             & Norm. ($10^{-3}$)$^j$ 		   & 	-			  		& 2.5$_{-0.1}^{+0.2}$ 	& 		-	  		& 	-					 \\
\\
{\sc rdblur}    	    & Betor10 $^k$ 				   & 	-				 	& -2.63$_{-0.05}^{+0.05}$ & -2.30$_{-0.07}^{+0.06}$ & -2.1$_{-0.1}^{+0.1}$	 \\
 		    	    & R$_{in}$ (R$_g$)$^l$ 		   & 	-					& 32$_{-7}^{+10}$  & 7$_{-1}^{+2}$	 & 7$_{-1}^{+6}$ 			 \\
 		    	    & R$_{out}$ (R$_g$)$^m$ 		   & 	-		  			& $10^4$ 				& $10^4$		  		 & $10^4$		 			 \\
 		    	    & Inclination (degree)$^n$ 		   & 	-			  		& 41$_{-3}^{+4}$	& 35.4$_{-0.9}^{+0.7}$  & 35$_{-3}^{+3}$ 		 \\
\\
{\sc rfxconv} 	    & Fe/solar$^o$ 				   & 	-			 		& 		-			&  =1 (fixed) 			& -				  		\\
 		   	    & Log$\xi$$^p$ 	   & 	-				 	& 		-			& $2.77_{-0.05}^{+0.05}$ 	& -  	 					 \\
 		   	    & Rel$_{refl}$ (sr)$^q$ 			   & 	-				 	& 		-			& $0.22_{-0.03}^{+0.05}$ 	& - 	 					\\
\\
{\sc highecut} 	    & CutoffE$^r$ 				   & 	-			 		& 		-			& 		-		 	& =0.2 (fixed) 				  \\
		 	    & FoldE$^s$ 				   & 	-			 		& 		-			& 		-		 	& =2.7$\times$ kT$_{e}$ 			  \\
{\sc reflionx} 	    & Fe/solar$^o$ 				   & 	-			 		& 		-			& 		-		 	& =1 (fixed) 				  \\
 		   	    & Log$\xi$$^p$ 	   & 	-				 	& 		-			& 		-		 	& $2.94_{-0.10}^{+0.09}$ 	\\
 		   	    & Norm. ($10^{-6}$)	$^t$ 	   	   & 	-				 	& 		-			& 		-		 	& $12.2_{-0.4}^{+0.5}$ 	\\

\hline
		 	    &$\chi^2/dof$$^u$				   &2126.22/1674			& 2091.52/1675		&2132.25/1677			& 2142.65/1677			\\
\end{tabular} 
\end{minipage}}
\end{center}
\begin{flushleft} $^a$ Neutral column density; $^b$ Blackbody temperature; $^c$ Normalization of the {\sc bbody} in unity of $L_{39}/D_{10kpc}^2$, where $L_{39}$ is the luminosity in unity of 10$^{39}$ erg s$^{-1}$ and $D_{10kpc}$ is the distance in unity of 10 kpc; $^d$ Electrons temperature of the corona; $^e$ Photon index;  $^f$ Seed photons temperature; $^g$ Normalization of {\sc nthcomp}; $^h$ Energy of the line; $^i$ Broadening of the line; $^j$ Normalization of the line; $^k$ Power law dependence of emissivity; $^l$ Inner radius in units of gravitational radii R$_g$; $^m$ Outer radius in units of gravitational radii R$_g$; $^n$ Inclination angle of the binary system; $^o$ Ratio of Iron and Hydrogen abundance; $^p$ Logarithm of the ionization parameter, with the latter expressed in units of erg cm s$^{-1}$; $^q$ Solid angle ($\Omega/2\pi$) subtended by the reflector as seen from the corona; $^r$ high energy cut-off of the reflection component, set as 2.7 times the electron temperature of the comptonisation model; $^s$ folding energy of the cut-off; $^t$ Normalization of the {\sc reflionx} component; $^u$ reduced $\chi^2$; \\
\end{flushleft}
\end{table*}
We find a neutral column density (N$_{H}$) of $\sim1.6\times10^{22}$ cm$^{-2}$, using the abundances of \citet{anders89}. At soft energies, a blackbody component with a temperature (kT$_{bb}$) of $\sim0.6$ keV is observed and it can be associated to a hot accretion disc as the inferred emission radius is $28\pm2$ km, which allows us to exclude emission from the surface of a 1.4 M$_{\odot}$ NS. The comptonizing component can be instead described by an electron temperature (kT$_e$) of $\sim2$ keV, a seed photons temperature (kT$_0$) of $\sim0.2$ keV and a photon index of 1.4. This corona probably originates in the inner regions of the disc or more likely in the boundary layer.  
{This model provides a poor fit ($\chi^2/dof$=4881.49/1688; Fig.~\ref{GX_3+1_continuum}), although we note that it should be the best choice for the continuum. Indeed other similar spectral continuum models, for example as {\sc blackbody+powerlaw} or a cold {\sc diskbb} \citep{mitsuda84} plus a hot {\sc blackbody}, were tried but they gave significantly worse statistical results ($\chi^2/dof$=112812.0/1700 and $\chi^2/dof$=5063.76/1690, respectively).} However, the poor fit of the {\sc bb+nthcomp} model is not surprising because residuals from the fit are observed at 2.6, 3.3, 3.9 and 6.4 keV, due to not modelled emission lines. 
We also notice the presence of an excess at energies higher than 15 keV. In order to improve the fit, we firstly introduce four {\sc gaussian} components letting the broadening ($\sigma$) to vary freely amongst the lines. An improvement in the fit is found ($\chi^2/dof$=2261/1676) and the measured centroid energy lines is 2.71, 3.33, 3.97 and 6.66 keV. These lines can thus be associated with S XVI, Ar XVIII, Ca XX (or XIX) and Fe XXV. The emission features are broad, with widths comprise between $\sim$0.1 and 0.3 keV (Fig.~\ref{GX_3+1_spectra}, top-left). This confirms the results of \citet{piraino12}, although we note that we have been able to identify also the S XVI line, which is only marginally seen in the spectral residuals of their work but statistically significant in our ($\Delta\chi^2$=71 for 2 additional dof). Furthermore, we add that the {\sc nthcomp} parameters are now changed: the photon index, the electron temperature and the seed photons temperature are increased to 2.7, 3 keV and 1.2 keV, respectively (see Table~\ref{table_continuum_comptb}).

The presence of an excess in the residuals from the fit at energies lower than 15 keV was not affected by the introduction of the lines. Therefore this may be the evidence that it is produced by an additional spectral component that we tentatively model with a {\sc powerlaw}. Hard powerlaws are not unusual in accreting NS (e.g. \citealt{disalvo00,disalvo01,iaria01}; \citealt{iaria01b}b; \citealt{disalvo04,paizis06,piraino07,migliari07}), although it has never been clearly observed in the spectra of GX 3+1.
The introduction of this component significantly improves the fit ($\chi^2/dof$=2126.22/1674) and, hereafter, it will be considered in all the spectral fits as its presence is always statistically significant ($\Delta\chi^2>70$ for 2 additional degree of freedom). We find that the photon index is very steep ($4.7\pm0.5$) and it is not able to completely flatten the high energy excess. We note however that a fit, with a slightly worse statistical significance ($\Delta\chi^2\sim10$), can also be found, when the photon index is consistent with $\sim3.4$. Although the former is statistically better, the photon index of the latter may be more likely.

We subsequently substitute the four {\sc gaussian} models with relativistic broadened emission features ({\sc diskline} in {\sc xspec}), in order to test the goodness of the relativistic reflection interpretation. For all the lines, we kept tied the inner disc radius (R$_{in}$), the outer disc radius (R$_{out}$), the inclination angle and the emissivity index of the illuminating source ($betor10$). We remark that the spectral fit is insensitive to the outer disc radius parameter, hence we left it frozen to the value of $10^4$ gravitational radii (R$_g=2GM_*/c^2$, where $M_*$ is the NS mass). These models provide a significant improvement if compared with the simple {\sc gaussian} models ($\chi^2/dof=2091.52/1675$) as would be expected if the lines are produced as a result of reflection (see Section~\ref{intro}). 
We found that the energy of the lines are consistent with those previously determined with the gaussian models and the relativistic parameters indicate that the system is inclined with an angle of $\sim 41$\textdegree. In addition, our results suggest that the inner accretion disc is possibly truncated away from the innermost stable circular orbit (R$_{in}\sim31$ R$_g$), confirming the findings of \citet{piraino12}. 
However, we propose that for a truncation radius of $\sim$30 R$_g$, we would expect an inner disc temperature of $\sim0.3$ keV which significantly differs from the 0.5 keV temperature found by the spectral analysis and, in particular, to its corresponding emission radius of $22\pm2$ km, (assuming a distance of 6.1 kpc), i.e. $\sim$11 R$_g$. We note how this result may be affected by the simplicity of the spectral model and it suggests that the adopted reflection model cannot provide a coherent picture for the underlying physics of the accretion processes in GX 3+1.
Therefore a self-consistent model appears to be the best choice for the spectral analysis of the source.

\subsection{Self-consistent models}
\label{selfcons}
The existence of four relativistic emission features suggests that they are produced by reflection of hard photons from the surface of the accretion disc. This makes GX 3+1 one of the most intriguing sources to be studied adopting self-consistent models which describe the reflection continuum component and the discrete features in accordance with the chemical composition of the accretion disc. To accomplish this study, we make use of a commonly used self-consistent model: the {\sc rfxconv} convolution model \citep{kolehmainen11}.
{\sc rfxconv} depends on five parameters: the relative reflection normalization rel$_{refl}$, the redshift of the source, the iron abundance relative to Solar ($Fe/Solar$), the cosine of the inclination angle and the ionization parameter ($Log \xi$) of the disc reflecting surface. We convolve this model with the {\sc nthcomp} and the {\sc powerlaw} models, as they are the continuum source of hard photons of the spectrum. We further convolve the {\sc rfxconv} component with a relativistic kernel ({\sc rdblur} in {\sc xspec}) to take into account relativistic distortion due to a rotating disc close to a compact object. As {\sc rfxconv} does not take into account the presence of Ar XVIII and Ca XIX features but only the Fe and S lines (which are more abundant), we include two {\sc diskline} components for them. In particular, we link the parameters of {\sc rdblur} to those of the {\sc diskline}. {We note that the Iron line is the strongest emission feature in the spectra and it drives the spectral parameters of the other lines. However, when the lines are left free to vary independently, the spectral parameters converge towards values consistent within the errors. It suggests that all the lines have a common origin.} The best fit with this model does not introduce any significant statistical improvement ($\chi^2/dof=$2132.25/1677, see Table~\ref{table_continuum_comptb}--column $\#$3 and figure~\ref{GX_3+1_spectra}-\textit{bottom-left}) with respect to the {\sc diskline} model but allows us to give a more physical interpretation of the spectrum of GX 3+1. We find that the solid angle ($\Omega/2\pi$) subtended by the reflector as seen from the corona is $0.22\pm0.5$ steradians, confirming the expectations of \citet{dove97}, which estimated a solid angle of 0.3 for a spherical geometry. We find also that the logarithm of the ionization parameter of the disc is $\sim2.8$, which can widely explain the ionization state of the four emission lines observed in the spectrum. The inclination angle of the system is found to be consistent with 35\textdegree. The iron abundance is compatible with the solar value (assumed to be 1) as the fit was almost insensitive to this parameter and, for this reason, we kept it fixed to unity. 
However the broadband continuum, with the introduction of this self-consistent model, is significantly changed: the photon index of the {\sc nthcomp} is decreased to $\sim1.8$ as well as the electron temperature which is $\sim 2.3$ keV. In addition, the most significant change is seen in the {\sc powerlaw} photon index and seed photons temperature which are decreased towards $\sim3.4$ and $\sim0.7$ keV, respectively. The photon index of the {\sc powerlaw} is now less extreme than that found in the previous section ($>4$) and it is able to fit satisfactorily the high energy excess. This result suggests that the continuum reflection component is required to model correctly the broadband continuum.
In addition, the inferred inner disc radius is consistent with 7 R$_g$ and hence largely discrepant in comparison to the simple {\sc diskline} model. This latter result is definitely more consistent with the temperature of the disc (i.e. $\sim$0.5--0.6 keV) and with its corresponding emission radius of $\sim15\pm2$ km, i.e. $\sim$7 R$_g$.
Furthermore, we note that the seed photons temperature and the blackbody temperature are very similar, suggesting that the inner region of the disc is the source of seed photons for comptonization and that, more likely, the optically thick corona is the boundary layer.

To validate our interpretation, we test another widely used self-consistent model: the {\sc reflionx} model \citep{ross05}. {The powerlaw photon index of the {\sc reflionx} component is tied to the photon index of the {\sc nthcomp} model, which appears to be the dominant source of hard photons (see Figure~\ref{GX_3+1_spectra}).} We then convolve the {\sc reflionx} model with a relativistic kernel ({\sc rdblur}, and also with a {\sc highecut} model in order to limit the spectral extension of the {\sc reflionx} model and mimic the roll-over of the comptonization component. The folding energy is fixed at 0.2 keV while the high energy cut-off is linked to the electron temperature of the {\sc nthcomp} model. The parameters of the {\sc rdblur} are then linked to those of the {\sc diskline} models used for the Ar XVIII and Ca XIX lines.

Not surprisingly, the best fit with this model is statistically consistent with the previous one ($\chi^2/dof$=2142.65/1677, see Table~\ref{table_continuum_comptb}, column $\#4$ and figure~\ref{GX_3+1_spectra}, \textit{bottom-right}), as well as the spectral findings. We note that the reflection component is produced by chemical elements with a logarithm of the ionization of $\sim$2.9, reasonably consistent (within the uncertainties) with the {\sc rfxconv} estimate. The inclination angle of the system is estimated again to be 35\textdegree. The relativistic parameters are also in line with the previous ones, and we find an inner disc radius roughly consistent with the innermost stable circular orbit. This confirms that the {\sc rfxconv} and {\sc reflionx} models give results consistent with each other and are clearly reproducing the same type of reflection process and geometry of the accretion flow, reinforcing our interpretations and conclusions.  

It is important to mention that the self-consistent reflection models that we used in this work do not take into account the Ar and Ca lines which appear to be important in the spectrum of GX 3+1. For this reason, in order to further investigate the nature of the Ar and Ca lines, we also tried the \textsc{relxill} model (\citealt{garcia14}, i.e. the relativistic version of the \textsc{xillver} model); this is currently the only self-consistent reflection model including Ar and Ca transitions and is, therefore, the best model to fit these low energy lines. We tested the latest public version of this model (v0.2g) where Ar and Ca transitions are included but it is not possible to independently vary their element abundances. The residuals, with respect to the best-fit, still point to an unaccounted excess at the energies of the resonant lines, suggesting a possible overabundance of these elements. We find, from the {\sc diskline} models, that the ratio of the equivalent widths of the Ar (Ca) line with respect to that of the Fe is $\sim$10 (13)$\%$. Using the simple calculations showed in \citet{disalvo15}, we might expect an overabundance of these elements of
$\sim$4 times the solar abundance. We note only that, in the GX 3+1 spectrum, they are highly statistically significant and, in addition, these lines are very often found in \textit{XMM-Newton} spectra (and perhaps also in \textit{Suzaku} spectra, see \citealt{disalvo15}) of bright LMXBs. In conclusion, at this moment we cannot infer any unambiguous indication about the abundance of these elements, which need to be further investigated with suitable reflection models.

Finally, we discuss the importance of the {\sc powerlaw} component: we verified that, without such a component, the fits with self-consistent models would provide unphysical results. In particular, the {\sc blackbody} temperature would have been decreased toward lower temperatures ($\sim$0.15 keV) although R$_{in}$ was still close to 6 R$_g$. Such a low temperature could only be consistent with a highly truncated accretion disc, which does not seem the case for GX 3+1.
{For completeness, we mention that we also tried an alternative model of the continuum in which the seed photons for comptonization were considered as originating from an accretion disc and with a blackbody component associated to a hot boundary layer/NS surface emission (``Western'' model, e.g. \citealt{white88}). The statistics of this model is marginally worse ($\chi^2/$dof$=2157.16/1677$) than the continuum model adopted above, but the results regarding the reflection parameters and the significance of the hard powerlaw are totally in line with those previously found. The only difference is related to the seed photons temperature ($\sim0.5$ keV) and the blackbody temperature ($\sim0.9$ keV) which are, however, simply reversed if compared with the previous ones.}

\section{Discussion and Conclusions} 
\label{discussion}

We analysed a \textit{XMM-Newton} observation, taken in timing mode, with the aim to investigate the reflection properties of the persistent accreting NS GX 3+1. In addition, we also made use of all the available \textit{INTEGRAL} observations collected from January to December, 2010. The whole \textit{INTEGRAL} data were stacked together to improve the signal-to-noise ratio. We have been able to study the X-ray emission from the source in the energy range 0.9--50 keV. 
GX 3+1 is a bright, accreting NS located in the direction of the Galactic center. We estimated an unabsorbed total flux of $\sim 7\times10^{-9}$ erg cm$^{-2}$ s$^{-1}$ and a luminosity of $\sim3\times10^{37}$ erg s$^{-1}$ (for a distance of 6.1 kpc). This corresponds to $\sim$20$\%$ of the Eddington luminosity.
The source is characterised by a complex spectral shape with asymmetrically broadened emission features which were associated to reflection of hard photons from the surface of the accretion disc. They were identified with the transitions of Iron, Calcium, Argon and Sulphur in H-like ionization state. We described their properties adopting either simple {\sc gaussian} or {\sc diskline} models, or self-consistent models. Our results with the {\sc diskline} models are consistent with the previous work of \citet{piraino12}, although the disc temperature is significantly lower in our work ($\sim$0.6 keV versus 0.9 keV) possibly due to the use of the RGS data. However, such models do not provide a global picture for the accretion mechanisms on GX 3+1. On the other hand, the spectral parameters obtained adopting self-consistent models are more coherent with each other and therefore, hereafter, we discuss the results of the self-consistent models as they provide a more physical interpretation of the X-ray emission from GX 3+1.

We modeled the continuum with a blackbody, more likely associated with the accretion disc, and a comptonizing component. The latter component is described by a cold ($\sim 2$ keV) electron population having a relatively flat spectrum ($\Gamma\sim2$) which decreases rapidly above 20 keV. This component is dominant at energies between 1.5--30 keV, corresponding to about 70$\%$ of the total flux emission, and the stability of this high energy component has been widely investigated \citep[e.g.][]{seifina12}. This makes feasible our choice to stack all the available \textit{INTEGRAL} observations. Such a component is probably produced in the transition layer around the NS and can partially cover the inner regions of the disc \citep{seifina12}. The accretion disc is instead relatively hot ($\sim$ 0.5-0.6 keV) and we inferred an emission radius of $15 \pm2$ km, suggesting it is close to the compact object, and with a temperature which is very similar to the seed photons temperature of the optically thick corona, strongly supporting the location of the corona close to the NS. In addition, we identified a steep powerlaw component ($\Gamma \sim3.4$) that does not show any indication of a high energy cut-off and that is predominant at energies higher than 30 keV. It is the first time that this component is observed in GX 3+1 and we found that it contributes to around 20$\%$ of the total flux. Hard powerlaws, also known as hard tails, are commonly seen during both low/hard and high/soft states of accreting black holes \citep{mcclintock06} and also in a number of accreting NS \citep[e.g.][]{disalvo00,disalvo01,damico01,iaria04,disalvo06,paizis06,dai07,tarana07,piraino07}. However the properties of the hard tails are different according to the spectral state in which they are observed. In particular, during the low/hard states, they dominate the broadband spectrum and peak above 100 keV, showing a cut-off at high energy which suggests an underlying thermal process \citep[e.g.][]{mcclintock06}. On the other hand, during the soft states, the hard tails are weak and carry only a few percent of the total flux emission and they do not present high energy cut-off. These can be explained as either being produced in a hybrid thermal/non thermal medium \citep{poutanen98} or by the bulk motion of accreting material close to the NS \citep[e.g.][]{titarchuk98}. An alternative scenario proposes a mechanisms of Comptonization of seed photons by relativistic, non thermal, electrons in an outflow \citep[e.g.][]{disalvo00} or synchrotron emission from a relativistic jet moving out from the system \citep{markoff01}. In the latter case, radio emission should be observed \citep[e.g.][]{migliari07,homan04}. According to the fact that GX 3+1 is more likely in a high/soft state and the powerlaw is weak and does not show any spectral cut-off, we suggest its hard tail is similar to those found during the high/soft state of the Galactic accreting systems. However, no radio emission has still been observed in GX 3+1 \citep{berendsen00} and a combined campaign of X-ray and radio observations would be required to clarify the accretion mechanisms.

In addition to the continuum components, GX 3+1 is characterized by four broad emission lines which are associated to relativistic reflection of hard photons from the surface of the disc. We inferred an inclination angle of system and a logarithm of the ionization parameter of the accretion disc consistent with 35\textdegree$ $ and 2.8, respectively. The inferred ionization parameter is in agreement with the H-like nature of the discrete features of the S, Ar, Ca and Fe that we found in the GX 3+1 spectrum.
The accretion disc is more likely responsible for the reflection component and the shapes of the emission lines are strongly dependent on the underlying continuum, therefore a good modelling of it can support more robust conclusions about the spectral interpretation. We argue that the study of the relativistic emission features of GX 3+1 with simple phenomenological models (e.g., {\sc diskline}) provides misleading interpretations of the accretion processes. In fact we found that the inner disc radius was consistent with $\sim 30$ R$_g$, suggesting that the disc is truncated away from the NS: however this is difficult to reconcile with a disc temperature of $\sim 0.5-0.6$ keV, as well as with the lack of X-ray pulsations probably reflecting the low magnetic field. The self-consistent models (both {\sc rfxconv} and {\sc reflionx}) instead point towards more physical results: the relativistic reflection is produced at a radius of $\sim10$ R$_g$ which is now highly consistent with the disc temperature of 0.5-0.6 keV (or $0.8-0.9$ keV, applying a color correction factor of 1.7). Furthermore, the disc and the reflection components contribute, together, to 10$\%$ of the total flux. We also finally note that the Ar and Ca lines in GX 3+1 are possibly over-abundant with respect to the solar one and further analyses and investigations are needed to better understand their nature.

We therefore conclude that the broadband spectral properties of GX 3+1 are complex and well described by a combination of a hot accretion disc and optically thick and thin coronae likely the sources of the hard photons responsible for the strong reflection component.

Although robust, our analysis was limited by the fact that the \textit{XMM-Newton} and \textit{INTEGRAL} datasets were not simultaneous. Therefore we strongly encourage further investigation of GX 3+1 with the support of broadband X-ray satellites such as \textit{Suzaku} and {\it NuSTAR} in order to provide a confirmation of the interpretations of the spectral results. 

\section*{Acknowledgements} 

We gratefully acknowledge the Sardinia Regional Government for the financial support (P. O. R. Sardegna F.S.E. Operational Programme of the Autonomous Region of Sardinia, European Social Fund 2007-2013 - Axis IV Human Resources, Objective l.3, Line of Activity l.3.1). This work was partially supported by the Regione Autonoma della Sardegna through POR-FSE Sardegna 2007-2013, L.R. 7/2007, Progetti di Ricerca di Base e Orientata, Project N. CRP-60529, and by the INAF/PRIN 2012-6. The High-Energy Astrophysics Group of Palermo acknowledges support from the Fondo Finalizzato alla Ricerca (FFR) 2012/13, project N. 2012-ATE-0390, founded by the University of Palermo.

\addcontentsline{toc}{section}{Bibliography}
\bibliographystyle{mn2e}
\bibliography{biblio}

\begin{thebibliography}{}

\bibitem[\protect\citeauthoryear{{Anders} \& {Grevesse}}{{Anders} \&
  {Grevesse}}{1989}]{anders89}
{Anders} E.,  {Grevesse} N.,  1989, \gca, 53, 197

\bibitem[\protect\citeauthoryear{{Arnaud}}{{Arnaud}}{1996}]{arnaud96}
{Arnaud} K.~A.,  1996, in {Jacoby}, G.~H. and {Barnes}, J., eds., Astronomical
  Data Analysis Software and Systems V. Vol.~101 of ASP Conf. Ser., San
  Francisco CA, {XSPEC: The First Ten Years}.
p.~17

\bibitem[\protect\citeauthoryear{{Berendsen}, {Fender}, {Kuulkers}, {Heise} \&
  {van der Klis}}{{Berendsen} et~al.}{2000}]{berendsen00}
{Berendsen} S.~G.~H.,  {Fender} R.,  {Kuulkers} E.,  {Heise} J.,    {van der
  Klis} M.,  2000, \mnras, 318, 599

\bibitem[\protect\citeauthoryear{{Bhattacharyya} \&
  {Strohmayer}}{{Bhattacharyya} \& {Strohmayer}}{2007}]{bhattacharyya07}
{Bhattacharyya} S.,  {Strohmayer} T.~E.,  2007, \apjl, 664, L103

\bibitem[\protect\citeauthoryear{{Bowyer}, {Byram}, {Chubb} \&
  {Friedman}}{{Bowyer} et~al.}{1965}]{bowyer65}
{Bowyer} S.,  {Byram} E.~T.,  {Chubb} T.~A.,    {Friedman} H.,  1965, Science,
  147, 394

\bibitem[\protect\citeauthoryear{{Cackett}, {Miller}, {Ballantyne}, {Barret},
  {Bhattacharyya}, {Boutelier}, {Miller}, {Strohmayer} \& {Wijnands}}{{Cackett}
  et~al.}{2010}]{cackett10}
{Cackett} E.~M.,  {Miller} J.~M.,  {Ballantyne} D.~R.,  {Barret} D.,
  {Bhattacharyya} S.,  {Boutelier} M.,  {Miller} M.~C.,  {Strohmayer} T.~E.,
  {Wijnands} R.,  2010, \apj, 720, 205

\bibitem[\protect\citeauthoryear{{Chenevez}, {Falanga}, {Brandt}, {Farinelli},
  {Frontera}, {Goldwurm}, {in't Zand}, {Kuulkers} \& {Lund}}{{Chenevez}
  et~al.}{2006}]{chenevez06}
{Chenevez} J.,  {Falanga} M.,  {Brandt} S.,  {Farinelli} R.,  {Frontera} F.,
  {Goldwurm} A.,  {in't Zand} J.~J.~M.,  {Kuulkers} E.,    {Lund} N.,  2006,
  \aap, 449, L5

\bibitem[\protect\citeauthoryear{{Courvoisier}, {Walter}, {Beckmann}, {Dean} \&
  {Dubath}}{{Courvoisier} et~al.}{2003}]{courvoisier03}
{Courvoisier} T.~J.-L.,  {Walter} R.,  {Beckmann} V.,  {Dean} A.~J.,
  {Dubath} P. e.~a.,  2003, \aap, 411, L53

\bibitem[\protect\citeauthoryear{{D'A{\`i}}, {di Salvo}, {Ballantyne}, {Iaria},
  {Robba}, {Papitto}, {Riggio}, {Burderi}, {Piraino}, {Santangelo}, {Matt},
  {Dov{\v c}iak} \& {Karas}}{{D'A{\`i}} et~al.}{2010}]{dai10}
{D'A{\`i}} A.,  {di Salvo} T.,  {Ballantyne} D.,  {Iaria} R.,  {Robba} N.~R.,
  {Papitto} A.,  {Riggio} A.,  {Burderi} L.,  {Piraino} S.,  {Santangelo} A.,
  {Matt} G.,  {Dov{\v c}iak} M.,    {Karas} V.,  2010, \aap, 516, A36

\bibitem[\protect\citeauthoryear{{D'A{\`i}}, {Iaria}, {Di Salvo}, {Matt} \&
  {Robba}}{{D'A{\`i}} et~al.}{2009}]{dai09}
{D'A{\`i}} A.,  {Iaria} R.,  {Di Salvo} T.,  {Matt} G.,    {Robba} N.~R.,
  2009, \apjl, 693, L1

\bibitem[\protect\citeauthoryear{{D'A{\'{\i}}}, {{\.Z}ycki}, {Di Salvo},
  {Iaria}, {Lavagetto} \& {Robba}}{{D'A{\'{\i}}} et~al.}{2007}]{dai07}
{D'A{\'{\i}}} A.,  {{\.Z}ycki} P.,  {Di Salvo} T.,  {Iaria} R.,  {Lavagetto}
  G.,    {Robba} N.~R.,  2007, \apj, 667, 411

\bibitem[\protect\citeauthoryear{{D'Amico}, {Heindl}, {Rothschild} \&
  {Gruber}}{{D'Amico} et~al.}{2001}]{damico01}
{D'Amico} F.,  {Heindl} W.~A.,  {Rothschild} R.~E.,    {Gruber} D.~E.,  2001,
  \apjl, 547, L147

\bibitem[\protect\citeauthoryear{{den Hartog}, {in't Zand}, {Kuulkers},
  {Cornelisse}, {Heise}, {Bazzano}, {Cocchi}, {Natalucci} \& {Ubertini}}{{den
  Hartog} et~al.}{2003}]{denHartog03}
{den Hartog} P.~R.,  {in't Zand} J.~J.~M.,  {Kuulkers} E.,  {Cornelisse} R.,
  {Heise} J.,  {Bazzano} A.,  {Cocchi} M.,  {Natalucci} L.,    {Ubertini} P.,
  2003, \aap, 400, 633

\bibitem[\protect\citeauthoryear{{di Salvo}, {D'A{\'{\i}}}, {Iaria}, {Burderi},
  {Dov{\v c}iak}, {Karas}, {Matt}, {Papitto}, {Piraino}, {Riggio}, {Robba} \&
  {Santangelo}}{{di Salvo} et~al.}{2009}]{disalvo09}
{di Salvo} T.,  {D'A{\'{\i}}} A.,  {Iaria} R.,  {Burderi} L.,  {Dov{\v c}iak}
  M.,  {Karas} V.,  {Matt} G.,  {Papitto} A.,  {Piraino} S.,  {Riggio} A.,
  {Robba} N.~R.,    {Santangelo} A.,  2009, \mnras, 398, 2022

\bibitem[\protect\citeauthoryear{{Di Salvo}, {Goldoni}, {Stella}, {van der
  Klis}, {Bazzano}, {Burderi}, {Farinelli}, {Frontera}, {Israel}, {M{\'e}ndez},
  {Mirabel}, {Robba}, {Sizun}, {Ubertini} \& {Lewin}}{{Di Salvo}
  et~al.}{2006}]{disalvo06}
{Di Salvo} T.,  {Goldoni} P.,  {Stella} L.,  {van der Klis} M.,  {Bazzano} A.,
  {Burderi} L.,  {Farinelli} R.,  {Frontera} F.,  {Israel} G.~L.,  {M{\'e}ndez}
  M.,  {Mirabel} I.~F.,  {Robba} N.~R.,  {Sizun} P.,  {Ubertini} P.,    {Lewin}
  W.~H.~G.,  2006, \apjl, 649, L91

\bibitem[\protect\citeauthoryear{{Di Salvo}, {Iaria}, {Matranga}, {Burderi},
  {D'Ai}, {Egron}, {Papitto}, {Riggio}, {Robba} \& {Ueda}}{{Di Salvo}
  et~al.}{2015}]{disalvo15}
{Di Salvo} T.,  {Iaria} R.,  {Matranga} M.,  {Burderi} L.,  {D'Ai} A.,  {Egron}
  E.,  {Papitto} A.,  {Riggio} A.,  {Robba} N.~R.,    {Ueda} Y.,  2015, ArXiv
  e-prints

\bibitem[\protect\citeauthoryear{{Di Salvo}, {Robba}, {Iaria}, {Stella},
  {Burderi} \& {Israel}}{{Di Salvo} et~al.}{2001}]{disalvo01}
{Di Salvo} T.,  {Robba} N.~R.,  {Iaria} R.,  {Stella} L.,  {Burderi} L.,
  {Israel} G.~L.,  2001, \apj, 554, 49

\bibitem[\protect\citeauthoryear{{di Salvo}, {Santangelo} \& {Segreto}}{{di
  Salvo} et~al.}{2004}]{disalvo04}
{di Salvo} T.,  {Santangelo} A.,    {Segreto} A.,  2004, Nuclear Physics B
  Proceedings Supplements, 132, 446

\bibitem[\protect\citeauthoryear{{Di Salvo}, {Stella}, {Robba}, {van der Klis},
  {Burderi}, {Israel}, {Homan}, {Campana}, {Frontera} \& {Parmar}}{{Di Salvo}
  et~al.}{2000}]{disalvo00}
{Di Salvo} T.,  {Stella} L.,  {Robba} N.~R.,  {van der Klis} M.,  {Burderi} L.,
   {Israel} G.~L.,  {Homan} J.,  {Campana} S.,  {Frontera} F.,    {Parmar}
  A.~N.,  2000, \apjl, 544, L119

\bibitem[\protect\citeauthoryear{{Dove}, {Wilms}, {Maisack} \&
  {Begelman}}{{Dove} et~al.}{1997}]{dove97}
{Dove} J.~B.,  {Wilms} J.,  {Maisack} M.,    {Begelman} M.~C.,  1997, \apj,
  487, 759

\bibitem[\protect\citeauthoryear{{Egron}, {Di Salvo}, {Motta}, {Burderi},
  {Papitto}, {Duro}, {D'A{\`i}}, {Riggio}, {Belloni}, {Iaria}, {Robba},
  {Piraino} \& {Santangelo}}{{Egron} et~al.}{2013}]{egron13}
{Egron} E.,  {Di Salvo} T.,  {Motta} S.,  {Burderi} L.,  {Papitto} A.,  {Duro}
  R.,  {D'A{\`i}} A.,  {Riggio} A.,  {Belloni} T.,  {Iaria} R.,  {Robba} N.~R.,
   {Piraino} S.,    {Santangelo} A.,  2013, \aap, 550, A5

\bibitem[\protect\citeauthoryear{{Fabian}, {Ballantyne}, {Merloni}, {Vaughan},
  {Iwasawa} \& {Boller}}{{Fabian} et~al.}{2002}]{fabian02}
{Fabian} A.~C.,  {Ballantyne} D.~R.,  {Merloni} A.,  {Vaughan} S.,  {Iwasawa}
  K.,    {Boller} T.,  2002, \mnras, 331, L35

\bibitem[\protect\citeauthoryear{{Fabian}, {Miniutti}, {Iwasawa} \&
  {Ross}}{{Fabian} et~al.}{2005}]{fabian05}
{Fabian} A.~C.,  {Miniutti} G.,  {Iwasawa} K.,    {Ross} R.~R.,  2005, \mnras,
  361, 795

\bibitem[\protect\citeauthoryear{{Fabian}, {Rees}, {Stella} \&
  {White}}{{Fabian} et~al.}{1989}]{fabian89}
{Fabian} A.~C.,  {Rees} M.~J.,  {Stella} L.,    {White} N.~E.,  1989, \mnras,
  238, 729

\bibitem[\protect\citeauthoryear{{Garc{\'{\i}}a} J. \&~{Dauser}, {Lohfink},
  {Kallman}, {Steiner}, {McClintock}, {Brenneman}, {Wilms}, {Eikmann},
  {Reynolds} \& {Tombesi}}{{Garc{\'{\i}}a} et~al.}{2014}]{garcia14}
{Garc{\'{\i}}a} J. \&~{Dauser} T.,  {Lohfink} A.,  {Kallman} T.~R.,  {Steiner}
  J.~F.,  {McClintock} J.~E.,  {Brenneman} L.,  {Wilms} J.,  {Eikmann} W.,
  {Reynolds} C.~S.,    {Tombesi} F.,  2014, \apj, 782, 76

\bibitem[\protect\citeauthoryear{{George} \& {Fabian}}{{George} \&
  {Fabian}}{1991}]{george91}
{George} I.~M.,  {Fabian} A.~C.,  1991, \mnras, 249, 352

\bibitem[\protect\citeauthoryear{{Hasinger} \& {van der Klis}}{{Hasinger} \&
  {van der Klis}}{1989}]{hasinger89}
{Hasinger} G.,  {van der Klis} M.,  1989, \aap, 225, 79

\bibitem[\protect\citeauthoryear{{Homan}, {Wijnands}, {Rupen}, {Fender},
  {Hjellming}, {di Salvo} \& {van der Klis}}{{Homan} et~al.}{2004}]{homan04}
{Homan} J.,  {Wijnands} R.,  {Rupen} M.~P.,  {Fender} R.,  {Hjellming} R.~M.,
  {di Salvo} T.,    {van der Klis} M.,  2004, \aap, 418, 255

\bibitem[\protect\citeauthoryear{{Iaria}, {Burderi}, {Di Salvo}, {La Barbera}
  \& {Robba}}{{Iaria} et~al.}{2001}]{iaria01b}
{Iaria} R.,  {Burderi} L.,  {Di Salvo} T.,  {La Barbera} A.,    {Robba} N.~R.,
  2001, \apj, 547, 412

\bibitem[\protect\citeauthoryear{{Iaria}, {D'A{\'{\i}}}, {di Salvo}, {Robba},
  {Riggio}, {Papitto} \& {Burderi}}{{Iaria} et~al.}{2009}]{iaria09}
{Iaria} R.,  {D'A{\'{\i}}} A.,  {di Salvo} T.,  {Robba} N.~R.,  {Riggio} A.,
  {Papitto} A.,    {Burderi} L.,  2009, \aap, 505, 1143

\bibitem[\protect\citeauthoryear{{Iaria}, {Di Salvo}, {Burderi} \&
  {Robba}}{{Iaria} et~al.}{2001}]{iaria01}
{Iaria} R.,  {Di Salvo} T.,  {Burderi} L.,    {Robba} N.~R.,  2001, \apj, 548,
  883

\bibitem[\protect\citeauthoryear{{Iaria}, {Di Salvo}, {Robba}, {Burderi},
  {Stella}, {Frontera} \& {van der Klis}}{{Iaria} et~al.}{2004}]{iaria04}
{Iaria} R.,  {Di Salvo} T.,  {Robba} N.~R.,  {Burderi} L.,  {Stella} L.,
  {Frontera} F.,    {van der Klis} M.,  2004, \apj, 600, 358

\bibitem[\protect\citeauthoryear{{Kallman} \& {White}}{{Kallman} \&
  {White}}{1989}]{kallman89}
{Kallman} T.,  {White} N.~E.,  1989, \apj, 341, 955

\bibitem[\protect\citeauthoryear{{Kolehmainen}, {Done} \& {D{\'{\i}}az
  Trigo}}{{Kolehmainen} et~al.}{2011}]{kolehmainen11}
{Kolehmainen} M.,  {Done} C.,    {D{\'{\i}}az Trigo} M.,  2011, \mnras, 416,
  311

\bibitem[\protect\citeauthoryear{{Kuulkers}}{{Kuulkers}}{2002}]{kuulkers02}
{Kuulkers} E.,  2002, \aap, 383, L5

\bibitem[\protect\citeauthoryear{{Kuulkers} \& {van der Klis}}{{Kuulkers} \&
  {van der Klis}}{2000}]{kuulkers00}
{Kuulkers} E.,  {van der Klis} M.,  2000, \aap, 356, L45

\bibitem[\protect\citeauthoryear{{Lebrun}, {Leray}, {Lavocat}, {Cr{\'e}tolle},
  {Arqu{\`e}s} \& {Blondel}}{{Lebrun} et~al.}{2003}]{lebrun03}
{Lebrun} F.,  {Leray} J.~P.,  {Lavocat} P.,  {Cr{\'e}tolle} J.,  {Arqu{\`e}s}
  M.,    {Blondel} C. e.~a.,  2003, \aap, 411, L141

\bibitem[\protect\citeauthoryear{{Lewin}, {van Paradijs}, {Hasinger},
  {Penninx}, {Langmeier}, {van der Klis}, {Jansen}, {Basinska}, {Sztajno} \&
  {Trumper}}{{Lewin} et~al.}{1987}]{lewin87}
{Lewin} W.~H.~G.,  {van Paradijs} J.,  {Hasinger} G.,  {Penninx} W.~H.,
  {Langmeier} A.,  {van der Klis} M.,  {Jansen} F.,  {Basinska} E.~M.,
  {Sztajno} M.,    {Trumper} J.,  1987, \mnras, 226, 383

\bibitem[\protect\citeauthoryear{{Lund}, {Budtz-J{\o}rgensen}, {Westergaard},
  {Brandt} \& {Rasmussen}}{{Lund} et~al.}{2003}]{lund03}
{Lund} N.,  {Budtz-J{\o}rgensen} C.,  {Westergaard} N.~J.,  {Brandt} S.,
  {Rasmussen} I.~L. e.~a.,  2003, \aap, 411, L231

\bibitem[\protect\citeauthoryear{{Mainardi}, {Paizis}, {Farinelli}, {Kuulkers},
  {Rodriguez}, {Hannikainen}, {Savolainen}, {Piraino}, {Bazzano} \&
  {Santangelo}}{{Mainardi} et~al.}{2010}]{mainardi10}
{Mainardi} L.~I.,  {Paizis} A.,  {Farinelli} R.,  {Kuulkers} E.,  {Rodriguez}
  J.,  {Hannikainen} D.,  {Savolainen} P.,  {Piraino} S.,  {Bazzano} A.,
  {Santangelo} A.,  2010, \aap, 512, A57

\bibitem[\protect\citeauthoryear{{Makishima}, {Mitsuda} \& {Inoue}}{{Makishima}
  et~al.}{1983}]{makishima83}
{Makishima} K.,  {Mitsuda} K.,    {Inoue} e.~a.,  1983, \apj, 267, 310

\bibitem[\protect\citeauthoryear{{Markoff}, {Falcke} \& {Fender}}{{Markoff}
  et~al.}{2001}]{markoff01}
{Markoff} S.,  {Falcke} H.,    {Fender} R.,  2001, \aap, 372, L25

\bibitem[\protect\citeauthoryear{{Matt}}{{Matt}}{2006}]{matt06}
{Matt} G.,  2006, Astronomische Nachrichten, 327, 949

\bibitem[\protect\citeauthoryear{{McClintock} \& {Remillard}}{{McClintock} \&
  {Remillard}}{2006}]{mcclintock06}
{McClintock} J.~E.,  {Remillard} R.~A.,  2006, {in Compact stellar X-ray
  sources, ed. W. H. G. Levin and M. van der Klis}.
Cambridge: Cambridge University Press, p.~157

\bibitem[\protect\citeauthoryear{{Migliari}, {Miller-Jones}, {Fender}, {Homan},
  {Di Salvo}, {Rothschild}, {Rupen}, {Tomsick}, {Wijnands} \& {van der
  Klis}}{{Migliari} et~al.}{2007}]{migliari07}
{Migliari} S.,  {Miller-Jones} J.~C.~A.,  {Fender} R.~P.,  {Homan} J.,  {Di
  Salvo} T.,  {Rothschild} R.~E.,  {Rupen} M.~P.,  {Tomsick} J.~A.,  {Wijnands}
  R.,    {van der Klis} M.,  2007, \apj, 671, 706

\bibitem[\protect\citeauthoryear{{Miller}, {Ballantyne}, {Fabian} \&
  {Lewin}}{{Miller} et~al.}{2002}]{miller02a}
{Miller} J.~M.,  {Ballantyne} D.~R.,  {Fabian} A.~C.,    {Lewin} W.~H.~G.,
  2002, \mnras, 335, 865

\bibitem[\protect\citeauthoryear{{Miller}, {D'A{\`i}}, {Bautz},
  {Bhattacharyya}, {Burrows}, {Cackett}, {Fabian}, {Freyberg}, {Haberl},
  {Kennea}, {Nowak}, {Reis}, {Strohmayer} \& {Tsujimoto}}{{Miller}
  et~al.}{2010}]{miller10}
{Miller} J.~M.,  {D'A{\`i}} A.,  {Bautz} M.~W.,  {Bhattacharyya} S.,  {Burrows}
  D.~N.,  {Cackett} E.~M.,  {Fabian} A.~C.,  {Freyberg} M.~J.,  {Haberl} F.,
  {Kennea} J.,  {Nowak} M.~A.,  {Reis} R.~C.,  {Strohmayer} T.~E.,
  {Tsujimoto} M.,  2010, \apj, 724, 1441

\bibitem[\protect\citeauthoryear{{Mitsuda}, {Inoue}, {Koyama}, {Makishima},
  {Matsuoka}, {Ogawara}, {Suzuki}, {Tanaka}, {Shibazaki} \& {Hirano}}{{Mitsuda}
  et~al.}{1984}]{mitsuda84}
{Mitsuda} K.,  {Inoue} H.,  {Koyama} K.,  {Makishima} K.,  {Matsuoka} M.,
  {Ogawara} Y.,  {Suzuki} K.,  {Tanaka} Y.,  {Shibazaki} N.,    {Hirano} T.,
  1984, \pasj, 36, 741

\bibitem[\protect\citeauthoryear{{Ng}, {D{\'{\i}}az Trigo}, {Cadolle Bel} \&
  {Migliari}}{{Ng} et~al.}{2010}]{ng10}
{Ng} C.,  {D{\'{\i}}az Trigo} M.,  {Cadolle Bel} M.,    {Migliari} S.,  2010,
  \aap, 522, A96

\bibitem[\protect\citeauthoryear{{Paizis}, {Farinelli}, {Titarchuk},
  {Courvoisier}, {Bazzano}, {Beckmann}, {Frontera}, {Goldoni}, {Kuulkers},
  {Mereghetti}, {Rodriguez} \& {Vilhu}}{{Paizis} et~al.}{2006}]{paizis06}
{Paizis} A.,  {Farinelli} R.,  {Titarchuk} L.,  {Courvoisier} T.~J.-L.,
  {Bazzano} A.,  {Beckmann} V.,  {Frontera} F.,  {Goldoni} P.,  {Kuulkers} E.,
  {Mereghetti} S.,  {Rodriguez} J.,    {Vilhu} O.,  2006, \aap, 459, 187

\bibitem[\protect\citeauthoryear{{Pintore}, {Sanna}, {di Salvo}, {Guainazzi},
  {D'A{\`i}}, {Riggio}, {Burderi}, {Iaria} \& {Robba}}{{Pintore}
  et~al.}{2014}]{pintore14b}
{Pintore} F.,  {Sanna} A.,  {di Salvo} T.,  {Guainazzi} M.,  {D'A{\`i}} A.,
  {Riggio} A.,  {Burderi} L.,  {Iaria} R.,    {Robba} N.~R.,  2014, ArXiv
  e-prints

\bibitem[\protect\citeauthoryear{{Piraino}, {Santangelo}, {di Salvo}, {Kaaret},
  {Horns}, {Iaria} \& {Burderi}}{{Piraino} et~al.}{2007}]{piraino07}
{Piraino} S.,  {Santangelo} A.,  {di Salvo} T.,  {Kaaret} P.,  {Horns} D.,
  {Iaria} R.,    {Burderi} L.,  2007, \aap, 471, L17

\bibitem[\protect\citeauthoryear{{Piraino}, {Santangelo}, {Kaaret}, {M{\"u}ck},
  {D'A{\`i}}, {Di Salvo}, {Iaria}, {Robba}, {Burderi} \& {Egron}}{{Piraino}
  et~al.}{2012}]{piraino12}
{Piraino} S.,  {Santangelo} A.,  {Kaaret} P.,  {M{\"u}ck} B.,  {D'A{\`i}} A.,
  {Di Salvo} T.,  {Iaria} R.,  {Robba} N.,  {Burderi} L.,    {Egron} E.,  2012,
  \aap, 542, L27

\bibitem[\protect\citeauthoryear{{Poutanen} \& {Coppi}}{{Poutanen} \&
  {Coppi}}{1998}]{poutanen98}
{Poutanen} J.,  {Coppi} P.~S.,  1998, Physica Scripta Volume T, 77, 57

\bibitem[\protect\citeauthoryear{{Reis}, {Fabian}, {Ross}, {Miniutti}, {Miller}
  \& {Reynolds}}{{Reis} et~al.}{2008}]{reis08}
{Reis} R.~C.,  {Fabian} A.~C.,  {Ross} R.~R.,  {Miniutti} G.,  {Miller} J.~M.,
    {Reynolds} C.,  2008, \mnras, 387, 1489

\bibitem[\protect\citeauthoryear{{Reynolds} \& {Nowak}}{{Reynolds} \&
  {Nowak}}{2003}]{reynolds03}
{Reynolds} C.~S.,  {Nowak} M.~A.,  2003, \physrep, 377, 389

\bibitem[\protect\citeauthoryear{{Ross} \& {Fabian}}{{Ross} \&
  {Fabian}}{2005}]{ross05}
{Ross} R.~R.,  {Fabian} A.~C.,  2005, \mnras, 358, 211

\bibitem[\protect\citeauthoryear{{Ross} \& {Fabian}}{{Ross} \&
  {Fabian}}{2007}]{ross07}
{Ross} R.~R.,  {Fabian} A.~C.,  2007, \mnras, 381, 1697

\bibitem[\protect\citeauthoryear{{Ross}, {Fabian} \& {Young}}{{Ross}
  et~al.}{1999}]{ross99}
{Ross} R.~R.,  {Fabian} A.~C.,    {Young} A.~J.,  1999, \mnras, 306, 461

\bibitem[\protect\citeauthoryear{{Seifina} \& {Titarchuk}}{{Seifina} \&
  {Titarchuk}}{2012}]{seifina12}
{Seifina} E.,  {Titarchuk} L.,  2012, \apj, 747, 99

\bibitem[\protect\citeauthoryear{{Tanaka}, {Nandra}, {Fabian}, {Inoue},
  {Otani}, {Dotani}, {Hayashida}, {Iwasawa}, {Kii}, {Kunieda}, {Makino} \&
  {Matsuoka}}{{Tanaka} et~al.}{1995}]{tanaka95}
{Tanaka} Y.,  {Nandra} K.,  {Fabian} A.~C.,  {Inoue} H.,  {Otani} C.,  {Dotani}
  T.,  {Hayashida} K.,  {Iwasawa} K.,  {Kii} T.,  {Kunieda} H.,  {Makino} F.,
   {Matsuoka} M.,  1995, \nat, 375, 659

\bibitem[\protect\citeauthoryear{{Tarana}, {Bazzano}, {Ubertini} \&
  {Zdziarski}}{{Tarana} et~al.}{2007}]{tarana07}
{Tarana} A.,  {Bazzano} A.,  {Ubertini} P.,    {Zdziarski} A.~A.,  2007, \apj,
  654, 494

\bibitem[\protect\citeauthoryear{{Titarchuk}, {Laurent} \&
  {Shaposhnikov}}{{Titarchuk} et~al.}{2009}]{titarchuk09}
{Titarchuk} L.,  {Laurent} P.,    {Shaposhnikov} N.,  2009, \apj, 700, 1831

\bibitem[\protect\citeauthoryear{{Titarchuk} \& {Zannias}}{{Titarchuk} \&
  {Zannias}}{1998}]{titarchuk98}
{Titarchuk} L.,  {Zannias} T.,  1998, \apj, 493, 863

\bibitem[\protect\citeauthoryear{{Ubertini}, {Lebrun}, {Di Cocco}, {Bazzano},
  {Bird} \& {Broenstad}}{{Ubertini} et~al.}{2003}]{ubertini03}
{Ubertini} P.,  {Lebrun} F.,  {Di Cocco} G.,  {Bazzano} A.,  {Bird} A.~J.,
  {Broenstad} K. e.~a.,  2003, \aap, 411, L131

\bibitem[\protect\citeauthoryear{{van den Berg}, {Homan}, {Fridriksson} \&
  {Linares}}{{van den Berg} et~al.}{2014}]{vandenberg14}
{van den Berg} M.,  {Homan} J.,  {Fridriksson} J.~K.,    {Linares} M.,  2014,
  \apj, 793, 128

\bibitem[\protect\citeauthoryear{{Vrtilek}, {Soker} \& {Raymond}}{{Vrtilek}
  et~al.}{1993}]{vrtilek93}
{Vrtilek} S.~D.,  {Soker} N.,    {Raymond} J.~C.,  1993, \apj, 404, 696

\bibitem[\protect\citeauthoryear{{White} \& {Holt}}{{White} \&
  {Holt}}{1982}]{white82a}
{White} N.~E.,  {Holt} S.~S.,  1982, \apj, 257, 318

\bibitem[\protect\citeauthoryear{{White}, {Stella} \& {Parmar}}{{White}
  et~al.}{1988}]{white88}
{White} N.~E.,  {Stella} L.,    {Parmar} A.~N.,  1988, \apj, 324, 363

\bibitem[\protect\citeauthoryear{{Zdziarski}, {Johnson} \&
  {Magdziarz}}{{Zdziarski} et~al.}{1996}]{zdiarski96}
{Zdziarski} A.~A.,  {Johnson} W.~N.,    {Magdziarz} P.,  1996, \mnras, 283, 193

\end{thebibliography}

\end{document}